\documentstyle[11pt]{article}
\begin{document}

\vspace*{-2.5cm}
{\bf\hfill MKPH-T-95-33\\}

\begin{center}
ON THE COUPLING OF THE $\eta $ MESON TO THE NUCLEON
\end{center}

\begin{center}
M. Kirchbach
\end{center}
\begin{center}
Institut f\"ur Kernphysik, TH Darmstadt, D--64289 Darmstadt,\\
Germany\\
\end{center}
\begin{center}
L. Tiator
\end{center}
\begin{center}
Institut f\"ur Kernphysik, Universit\"at Mainz, D--55099 Mainz,\\
Germany\\
\end{center}

\begin{abstract}
The pseudoscalar and pseudovector $\eta N$ coupling constants
are calculated from an effective
vertex associated with the $a_0(980)\pi N$ triangle diagram.
The predicted values are in agreement with 
the ones concluded from
fitting $\eta $ photoproduction amplitudes.
In this context 
we stress the importance of the properties of
the scalar meson octet for $\eta $ meson physics.
\end{abstract}

\section{Introduction}

In contrast to the $\pi N$-interaction, little is known about the $\eta
N$-interaction and, consequently, about the $\eta NN$ vertex.
In the
case of pion scattering and pion photoproduction the $\pi NN$ coupling
is preferred to be pseudovector (PV), in accord with current algebra
results and chiral symmetry.  However, because the eta mass is so much
larger than the pion mass - leading to large SU(3) x SU(3) symmetry
breaking - and because of the $\eta-\eta'$ mixing there is no compelling reason
to select the PV rather than
the PS form for the $\eta NN$ vertex.

The uncertainty regarding the structure of the $\eta NN$ vertex
extends to the magnitude of the coupling constant.
This coupling constant $g_{\eta NN}^2/4\pi$ varies between $0$ and $7$
with the large couplings arising from fits of one boson exchange potentials.
Typical values obtained in
fits with OBEP potentials \cite{Bro90} can lie anywhere between 3 - 7.
However, including the $\eta$ yields only small effects in fitting the
$NN$ phase
shifts and, furthermore, provides an insignificant contribution to nuclear
binding at normal nuclear densities.
Furthermore such OBEP potentials use the eta as an effective meson to describe
effects of more elaborate two-meson correlations. This can be seen in the full
Bonn potential, where the eta coupling is below 1 and can be neglected in the 
calculations \cite{Ma89,Hol92}.
From SU(3) flavor symmetry all coupling constants between the meson octet
and the baryon octet are determined by one free parameter $\alpha$, giving
\begin{equation}
\frac{g_{\eta NN}^2}{4 \pi} = \frac{1}{3}(3-4\alpha)^2\;
\frac{g_{\pi NN}^2}{4 \pi}\;.
\end{equation}
The resulting values for the coupling constant lie
between $0.8$ and $1.9$ for commonly used values of
$\alpha$ between $0.6 - 0.65$ and depend on the F and D strengths chosen as the
two types of SU(3) octet meson-baryon couplings.
Other determinations of the $\eta NN$
coupling employ reactions involving the eta, such as
$\pi^- p \rightarrow \eta n$, and range from 
$g_{\eta NN}^2/{4 \pi} = 0.6 - 1.7$ \cite{Pen87}.
Smaller values are supported by $NN$ forward dispersion
relations \cite{Grein80} with
$g_{\eta NN}^2/{4 \pi}+g_{\eta' NN}^2/{4 \pi}\leq 1.0$.
There is some rather indirect evidence that also favors a small value for
$g_{\eta NN}$.
In Ref. \cite{Pie93}, Piekarewicz calculated the $\pi$--$\eta$
mixing amplitude in the hadronic model where the mixing was generated by
$\bar N N$ loops and thus driven by the proton--neutron mass difference.
To be in agreement with results from chiral perturbation theory the $\eta
NN$ coupling had to be constrained to the range 
$g_{\eta NN}^2/{4 \pi} = 0.32 - 0.53$.  In a very
different approach, Hatsuda \cite{Hat90} evaluated the proton matrix
element of the flavor singlet axial current in the large $N_C$ chiral
dynamics with an effective Lagrangian that included the $U_A$(1)
anomaly.  In this framework, the EMC data on the polarized proton
structure function (which have been used to determine the "strangeness
content" of the proton) can be related to the $\eta 'NN$ and the $\eta
NN$ coupling constants.  Again, his analysis prefers small values for
both coupling constants.  Nevertheless, from the above
discussion it seems clear that the $\eta NN$ coupling constant is much
smaller compared to the corresponding $\pi NN$ value of around 14.

In a recent analysis of $\eta$ photoproduction on the proton \cite{Tia94}
both nature and magnitude could be determined in a comparison of a dynamical
model with new high accuracy data from Mainz \cite{Kru95}. In this calculation
the resonance sector includes the $S_{11}(1535)$,
$P_{11}(1440)$ and $D_{13}(1520)$ states whose couplings are
fixed by
independent electromagnetic and hadronic reactions like $(\gamma,\pi)$,
$(\pi,\pi)$, $(\pi,\pi\pi)$ and $(\pi,\eta)$.
The nonresonant background is described by vector meson exchange contributions 
and s- and u-channel Born terms, where the $\eta NN$ coupling constant enters.
By comparison with the data on total and differential cross sections the 
couping constant was determined as  $g_{\eta NN}^2/4\pi \approx 0.4$
with a clear preference for a pseudoscalar type.

The aim of this paper is to calculate the pseudovector
as well as the pseudoscalar coupling constant of the $\eta $ meson 
to the nucleon.
In section 2 we analyze the different 
structures of the isosinglet and isotriplet 
axial vector nucleon currents on the quark level
to motivate the smallness of 
the pseudovector $\eta N$  coupling (subsequently denoted by 
$f_{\eta NN}$) relative to the corresponding pseudovector
$\pi N$ coupling (denoted by $f_{\pi NN}$).
In fact, in contrast to the isotriplet axial
current the isosinglet
axial vector current of the nucleon
does not contain any component formed of the isodoublet
$u$ and $d$ quarks of the first quark generation but
is determined exclusively by the 
isosinglet $c$ and $s$ quarks belonging to the
second quark generation.
The (pointlike) pseudovector $\eta N$ coupling
will be therefore exclusively determined by the presence of
strange/charmed  quarkonium
component both in the meson wave function
and the nucleon current
and expected to be rather small.
In this context vertex corrections
can acquire importance.
We here advocate the idea to treat the coupling of strange
quarkonium to the nucleon by means of triangular
vertices involving appropriate nonstrange mesons.

In section 3 we consider the long range 
$ \eta \to (a_0 \pi N)$ triangle diagram
as a model for the mixture of the pseudoscalar and pseudovector
$\eta NN$ vertices,
derive analytical expressions for $g_{\eta NN}$ and 
$f_{\eta NN}$ and fix in a natural way their relative sign.
The special role of the $a_0(980)\pi N$ triangle diagram
as the dominant one--loop mechanism for the $\eta N$ coupling 
is singled out by the circumstance that the $a_0(980)$ meson is
the lightest meson with a two particle decay channel
containing the $\eta $ particle \cite{PDG94}.
The contributions of heavier mesons such as
the isotriplet $a_2(1320)$ tensor meson with an $\eta\pi $ 
decay channel and the
isoscalar $f_0(1400)$, $f^\prime _2(1525)$ and $f_2(1720)$ tensor
mesons with $\eta \eta $ decay channels
will be left out of consideration
because of the short range character
of the corresponding triangle diagrams on the one side,
\footnote{The same argument applies to
the neglect of the $f_0(1590)\eta N$ triangular
vertex.}
and because of the comparatively small couplings of the tensor
mesons to the nucleon \cite{Ma89,Els87} on the other side.

In section 4 we show that the small value
of $g_{\eta NN}^2/4\pi \approx 0.4$,
as obtained from fits
of the photoproduction amplitude \cite{Tia94}
is well reproduced in terms of the $a_0(980)\pi N$ triangular 
$\eta N$ coupling if
complete dominance of the full $a_0$ decay width
by the $a_0(980)\to \eta + \pi $ decay channel is assumed
and use is made by the version of the
Bonn potential with the lowest value for the $a_0N$ coupling 
constant \cite{Ma89}.
The paper ends with a short summary.

\section{The couplings of the $\eta $ meson to the isoscalar
axial vector current of the nucleon }

Within the SU(3) flavor symmetry scheme
the neutral weak axial current of the nucleon 
$J_{\mu ,5}^0$ following from
the Glashow--Weinberg--Salam electroweak
gauge theory is given by the matrix element of the 
corresponding quark currents as
\begin{equation}
J_{\mu ,5}^0  = 
-{1\over 4}\langle N\mid \bar u\gamma_\mu\gamma_5 u -
\bar d\gamma_\mu\gamma_5  d - \bar s\gamma_\mu\gamma_5 s 
\mid N \rangle \, .
\end{equation}
This may be reduced to 
\begin{eqnarray}
J_{\mu ,5}^0 & = & -{1\over 2}\langle N\mid 
\left (
\begin{array}{rr}
\bar u &  \bar d \\ 
\end{array}
\right )
\gamma_\mu\gamma_5 {\tau_3\over 2}\, {u\choose d}
\mid N\rangle +
               {1\over 4}  <N\mid \overline{s}\gamma_\mu\gamma_5 s
\mid N> \nonumber\\
& = & -{g_A\over 2} \overline{u}(\vec{p}\, '\, )
\gamma_\mu\gamma_5 {\tau_3\over 2}u(\vec{p}\, ) 
+ {1\over 2}G_1^s \overline{u}(\vec{p}\, '\, )
\gamma_\mu\gamma_5u(\vec{p}\, )  \nonumber\\
& \equiv  & {1\over 2}(-J_{\mu ,5}\, (I=1) + J_{\mu ,5}\, (I=0)\,) .
\end{eqnarray}
Here $g_A$ is the weak isovector axial coupling constant,
$G_1^s$ denotes the weak isoscalar axial coupling
($g_A$= 1.25, $G_1^s =- 0.13 \pm 0.04 $ \cite{Ell93}), 
$u (\vec{p}\, )$ stands for the Dirac bi--spinor of the nucleon, whereas
$J_{\mu ,5}(I=1)$ and $J_{\mu ,5}(I=0)$ in turn denote
the isotriplet and isosinglet axial vector nucleon currents.
The matrix element of the isoscalar axial vector current
between the $\eta/\eta^\prime/\pi $
pseudoscalar mesons and  a $N\bar N$---state is defined in the
standard way as

\begin{eqnarray}
\langle N\bar N\mid J_{\mu ,5}(I=0)\mid \eta \rangle
 & = & if_\eta  m_\eta q_\mu  \, ,\nonumber\\
\langle N\bar N\mid J_{\mu ,5}(I=0)\mid \eta^\prime \rangle
 & = & if_{\eta^\prime}  m_{\eta^\prime} q_\mu  \, ,\nonumber\\
\langle N\bar N\mid J_{\mu ,5}(I=1)\mid \pi \rangle
 & = & if_{\pi}  m_{\pi} q_\mu  \, .
\end{eqnarray}
Here $m_n$ and $f_n$ 
denote in turn the mass and the \underline{dimensionless} coupling constant 
of the respective meson ($n=\eta ,\eta^\prime , \pi $ )
to the hadronic vacuum.

In the three flavor quark model, 
the wave functions of the low lying pseudoscalar
mesons are described 
as linear combinations of quark--antiquark ($q\overline{q}$) pairs. 
The physical 
singlet and scalar states within the pseudoscalar meson octet
corresponding to the $\eta^\prime$
and the $\eta $ mesons are moreover predicted to be mixed
according to
\begin{eqnarray}
\mid \eta  \rangle & = & \cos\theta_P 
{{\bar u u + \bar d d-2\bar s s}\over \sqrt{6}} - \sin\theta_P
{{\bar u u + \bar d d + \bar s s}\over \sqrt{3}}  \, , \nonumber\\
\mid \eta^\prime (958)\rangle & = &
\sin\theta_P
{{\bar u u + \bar d d-2\bar s s}\over \sqrt{6}} 
+ \cos \theta_P
{{\bar u u + \bar d d + \bar s s}\over \sqrt{3}}  \, ,
\end{eqnarray}
with the mixing angle $\theta_P = -10.1^\circ $ \cite{PDG94}
being determined from mass formulae.
The presence of a strange quarkonium component
in the pseudoscalar isoscalar mesons
is equivalent to  a violation of the Okubo--Zweig--Iizuka rule 
predicting the suppression of 
$\overline{s}s\to \overline{u}u / \overline{d}d$ transitions.

The evaluation of the matrix element of the isoscalar axial
vector current between the $\eta $ meson state and the
hadronic vacuum 
is based on the QCD suggestion \cite{Jaf89} that
a quark $q_i$ of flavor $i$ in the pseudoscalar mesons couples
only to the current $\bar q_i\gamma_\mu\gamma_5 q$
of the same flavor and that the coupling strength $\kappa $ is flavor
independent
\begin{equation}
{1\over 2}\langle \bar N N \mid \bar q_i \gamma_\mu\gamma_5 q_i
\mid  n \rangle 
 =  i\kappa\,  \alpha_n^j \,
\, m_n q_\mu \delta_{ij} \, . 
\end{equation}

Here, $\alpha_n^j$ denotes the weight of the $(\bar q_j q_j)$ quarkonium
in the wave function of the pseudoscalar meson 
($n=\eta, \eta^\prime ,\pi$ )
considered. 
Eq. (3) shows that in contrast to the
isoscalar vector current, 
the isoscalar axial vector current of the nucleon
contains no non--strange component.
Because of that the pointlike 
coupling of the isosinglet pseudoscalar mesons
to the corresponding nucleon current is realized only via their
strange quarkonium components, in which case one has
\begin{eqnarray}
\alpha^s_\eta & = &-{1\over \sqrt{6}} \cos\theta_P 
-{1\over {2\sqrt{3}}}\sin\theta_P\,, \\
\alpha^s_{\eta^\prime} & = & -{1\over \sqrt{6}} \sin\theta_P +
{1\over {2\sqrt{3}}} \cos\theta_P    \,,
\end{eqnarray}
while for the pion one has\\
\begin{equation}
\alpha_\pi = {1\over {\sqrt{2}}}\, . 
\end{equation}

Insertion of Eqs. (7-9) into (6) and a subsequent
comparison with (4) lead on one side to
\begin{equation}
f_\pi  = \kappa\alpha_\pi = {2\over 3} \, , 
\end{equation}
where we made use of the empirical value for the (dimensionless)
pion decay coupling constant $f_\pi = 92{\rm MeV}/m_\pi $.
On the other side, with that the coupling strength $\kappa $ is calculated
as $\kappa = 0.9428 $
and the values of 
$f_\eta $ and $f_{\eta^\prime}$ are completely determined by
\begin{eqnarray}
f_\eta & =& \kappa \alpha^s_\eta \,,\\
f_{\eta^\prime} & = & \kappa \alpha^s_{\eta^\prime}\,,
\end{eqnarray} respectively.
To get a rough understanding of the origin of
the pseudovector
$\eta N$, $ \eta^\prime\, N$ and $\pi N$ couplings
introduced via
the corresponding Lagrangians as
\begin{eqnarray}
{\cal L}_{\eta /\eta^\prime }(x) & = & 
{f_{(\eta /\eta^\prime )NN}\over m_{\eta /\eta^\prime }}
                            \bar \Psi (x)\gamma_\mu \gamma_5
                            \Psi(x)\partial^\mu\phi_{\eta /\eta^\prime}(x)\, , \\
{\cal L}_{\pi }(x) & = & 
{f_{ \pi NN}\over m_{\pi }}
                            \bar \Psi (x)\gamma_\mu \gamma_5
\vec \tau
                            \Psi(x)\cdot \partial^\mu\vec{\phi}_{\pi }(x)\, , 
\end{eqnarray}
it is quite instructive
to consider a ''toy'' model
in which universality of the axial currents of the
pseudoscalar meson is assumed for the moment (Fig.~1).
This would allow one to obtain the following parametrizations
\begin{eqnarray}
{g_{\eta NN}^{\rm toy}\over {2m_N}}= 
{f_{\eta NN}^{\rm toy}\over m_\eta}
 &=&  {G_1^s\over {2f_\eta m_\eta}}
\, = {0.1962\over m_\eta}\, , \\
{g_{\eta^\prime  NN}^{\rm toy}\over {2m_N}}= 
{  f_{\eta^\prime NN}^{\rm toy}\over m_\eta} &=&  
{ G_1^s \over {2f_{\eta^\prime} m_{\eta^\prime} }}\, = 
{{-0.1941}\over m_{\eta^\prime}},\\
{g_{\pi NN}^{\rm toy}\over {2m_N}}= 
{  f_{\pi NN}^{\rm toy}\over m_\pi } &=&  
{ g_A\over {2f_{\pi} m_{\pi} }}\, = {0.9375\over m_\pi}\, ,
\end{eqnarray}
where use has been made of the on--shell equivalence
between the pseudoscalar and pseudovector meson nucleon
couplings leading to the relation $f_n/m_n = g_{nNN}/2m_N$.
The usefulness of the 
''universality'' ansatz is best
demonstrated for the case
of the pion where the empirical value of 
$f_{\pi NN} = 1.0026$ as deduced
with a good accuracy from chiral symmetry constraints
is only few percent larger
than the one concluded from the
''universality '' arguments as 
$f_{\pi NN}^{\rm toy} = g_A/2 f_\pi = 0.9375 f_{\pi NN} $.
For the case of the charged axial vector current ''universality''
is equivalent to the Goldberger--Treiman (GT) relation and thus
to current conservation in the chiral limit of
a vanishing pion mass.
For the case of the isoscalar axial vector current, however, 
the ''toy'' model is less useful as it would suggest 
a GT--like relation between 
$G_1^s, f_\eta , f_{\eta NN}$ and $m_\eta$, which is unrealistic
in view of the axial anomaly problem.
Nontheless, the considerations given above are
instructive in a sense that they clearly 
illustrate the fundamental difference between the couplings of
isovector and isoscalar pseudoscalar mesons to the axial nucleon current.
Whereas the pseudovector $\eta /\eta^\prime N$
coupling relies on the strange component of the
axial vector current, its purely non strange component
is relevant for the pseudovector $\pi N$ coupling.
Eqs. (10-12) together with
Eqs. (15-17) lead to the following relations 
\begin{eqnarray}
{f_\eta\over f_\pi} &=& {\alpha_\eta \over \alpha_\pi}
=  \sqrt{2}\, (-{1\over \sqrt{6}}\cos\theta_P - 
{1\over {2\sqrt{3}}}\sin\theta_P)\, ,\\
r= {{f^{\rm toy}_{\eta NN}\, {m_\pi } } \over {f_{\pi NN}^{\rm toy}\, m_\eta }} 
& = &
{g_{\eta NN}\over g_{\pi NN}}
= {{G_1^s f_\pi \, m_\pi}\over {g_A f_\eta \,  m_\eta }} = - 0.0527\, .
\end{eqnarray}
Eq. (19) shows
that the $\eta N$ vertex appears suppressed
relative to the $\pi N$ vertex by at least one
order of magnitude.
For this reason we expect the much larger experimentally
observed $\eta N$ couplings $(r\approx 0.2)$ 
to be governed mainly by the effective
$a_0\pi N$ triangular vertex rather than by the
contact meson--current couplings considered in the ''toy'' model
above.

In the following section we consider
an effective $\eta NN$ vertex associated with the
$a_0\pi N$ triangle diagram (Fig. 2), which is
the dominant long range one loop mechanism
for the isoscalar axial nucleon coupling,
and calculate both the values  of $f_{\eta NN}$ and $g_{\eta NN}$
associated with this vertex.

\section{The $ a_0(980)\pi  N$ triangular vertex for
$g_{\eta NN}$ and $f_{\eta NN}$}

The $\pi a_0 N $ triangle diagram
is calculated using the following effective Lagrangians
for the $ a_0 \to \eta +\pi$ decay, the $\pi N$ and the $a_0 N$
couplings:

\begin{eqnarray}
{\cal L}_{a_0\eta \pi }(x) & = & f_{a_0\eta \pi } 
{{m_{a_0}^2 -m_\eta^2}\over m_\pi}\phi_\eta^\dagger (x) 
\vec{\phi_\pi}(x)\cdot
\vec{\phi}_{a_0}(x) \\
{\cal L}_{\pi NN}(x) & = & 
\frac{f_{\pi NN}}{m_\pi}\bar\psi(x)\gamma_\mu \gamma_5
\vec \tau\psi (x)\cdot \partial^\mu\vec {\phi}_\pi (x) , \\ 
{\cal L}_{a_0NN}(x)& = & g_{a_0 NN}\bar\psi (x)
\vec\tau\psi (x)\cdot\vec{\phi}_{a_0}(x) \, .
\end{eqnarray}
Here $f_{\pi NN}$ and $g_{a_0NN}$ 
in turn denote the pseudovector $\pi N$ 
and  the scalar  $a_0N$ coupling constants, for which we
adopt the values $f_{\pi NN}^2/4\pi = 0.08$ and 
$g_{a_0 NN}^2/4\,\pi=0.77$, respectively. These values are implied
by the relativistic Bonn one boson
exchange potential (OBEPQ) for the nucleon-nucleon interaction
\cite{Ma89}. 
To regularize the integral
in the triangle diagrams in Fig. 2 we introduce the same monopole
form factors at the $\pi$NN and $a_0$NN  vertices as established by
the Bonn potential model.
The following contribution to
the $\eta NN$ vertex is then obtained:
\begin{eqnarray}
g_{\eta NN}(q^2) & = &
\frac{3}{8\pi^2}{{m_{a_0}^2-m_\eta^2}\over m_\pi^2}
 f_{\pi NN}f_{a_0\eta \pi}g_{a_0 NN}\nonumber\\
& & \biggl\{   \int_0^1dx \ln
 \frac{{\cal Z}_1(m_\pi,\Lambda_{a_0}, x,q^2)
{\cal Z}_1(m_{a_0},\Lambda_\pi, x,q^2)}
{ {\cal Z}_1(m_\pi,m_{a_0}, x,q^2)
{\cal Z}_1(\Lambda_\pi,\Lambda_{a_0} , x,q^2)}   \nonumber\\
& +& \int_0^1\int_0^1 dy dx {{xc(x,y,1-y, q^2)}\over
{{\cal Z}_2(m_\pi,m_{a_0}, x,y, 1-y,q^2)}} \nonumber\\
&+ &  {1\over 2}\int_0^1\int_0^1 dy dx x \Bigl( \ln
{ { {\cal Z}_2 (m_\pi,\Lambda_{a_0}, x,1-y,q^2) } \over
{ {\cal Z}_2(m_\pi,m_{a_0}, x,1-y,q^2)} }\nonumber\\
& + &  \ln 
{ { {\cal Z}_2(m_{a_0},\Lambda_\pi, x,1-y,q^2)}\over
{ {\cal Z}_2(\Lambda_\pi,\Lambda_{a_0} , x,1-y,q^2)} } \Bigr) \biggr\}\,.
\end{eqnarray}
Here $\Lambda_\pi$ and $\Lambda_{a_0}$ are the cut-off parameters
in the monopole vertex factors, for which we use 
the values
1.05 GeV and 2.0 GeV, respectively.
The functions ${\cal Z}_1(m_1,m_2,x,q^2)$ and 
${\cal Z}_2(m_1,m_2, x, \bar y,q^2)$ are
defined as
\begin{eqnarray}
c(x, y,\bar y, q^2) & =& x\bar y (1+x\bar y )m_N^2 +
xy( x(y - \bar y) -{1\over 2})q^2 \, ,\nonumber\\
{\cal Z}_1(m_1,m_2,x,q^2) & = & xm_1^2 +(m_2^2-q^2)(1-x) 
+ (1-x)^2q^2\, , \nonumber\\
{\cal Z}_2(m_1,m_2, x,y,\bar y , q) & =& 
m_N^2x^2\bar y^2 +m_1^2(1-x) + (m_2^2-m_1^2)xy \nonumber\\
& + & x^2y(y-\bar y)q^2\, .
\end{eqnarray}
The $\eta N$ coupling constant is obtained by setting 
$q^2= m_\eta^2 $ in Eq. (23).
The corresponding expression for the pseudovector coupling reads:
\begin{eqnarray}
f_{\eta NN}(q^2) & = &
\frac{3}{8\pi^2}{{m_{a_0}^2-m_\eta^2}\over m_\pi^2}
 f_{\pi NN}f_{a_0\eta \pi}g_{a_0 NN} m_Nm_\eta \nonumber\\
& & \int_0^1\int_0^1 dy dx x^2 y  
\Bigl(Z^{-1}_2 (m_\pi,m_{a_0}, x,y, 1-y,q^2)\nonumber\\
& + &  
Z^{-1}_2(m_\pi,m_{a_0}, x,y, y-1,q^2)\Bigr)\,.
\end{eqnarray}

\section{Results and discussion}

Using for $f_{a_0\eta \pi}$ the value of 0.44 extracted 
from the experimental
decay width \cite{PDG94} when ascribing the total $a_0$
decay width to the $a_0\to \eta + \pi$ decay channel, 
we obtain
\begin{eqnarray}
g_{\eta NN} & = &2.03\, ,\qquad 
{g_{\eta NN}^2\over {4\pi}} =0.33\, ,\\
f_{\eta NN} & = & 0.58\, , \qquad {f_{\eta NN}^2\over {4\pi}} = 0.027\, .
\end{eqnarray}
These are the quantities which we shall
interpret as the values for the pseu\-do\-sca\-lar and 
pseu\-dovec\-tor
coupling constants, respectively.
The main sources of uncertainty in the parametrization of the
$\eta NN$ coupling constants by means of
the triangular $a_0(980)\pi N$ diagram are
associated with the $a_0(980) N$ coupling constant and
the $\Gamma (\eta \pi)/ \Gamma^{\rm tot}_{a_0}$
fraction.
The coupling constant $g_{a_0NN}$ varies between $3.11$ and $5.79$
depending on the $NN$ potential model version 
\cite{Ma89, Els87}. 
In view of the
$K\bar K$ mesonium structure of the
$a_0$ meson \cite{PDG94}
the $a_0 N$ coupling will be mainly governed
by the short range $K\bar K\Lambda $ intermediate configuration
and therefore expected to be small.
For this reason we favor in the present investigation
the versions of the Bonn potential 
with the lowest $g_{a_0 NN}$ values reported.
It should further be pointed out
that an
increase of $g_{\eta NN}$ and $f_{\eta NN}$
implied by a larger $g_{a_0 NN}$ value
can be compensated to a large amount by
the reduction of the $a_0(980)\to \eta +\pi $ partial width
from the 100\% used by us to a lower and more realistic value.
The size of the coupling constants obtained in the
present study can therefore be viewed as realistic.
Note that the pseudovector $\eta N$
effective coupling constant associated with the
$a_0\pi N$ triangle is about three times larger as compared to the
corresponding toy model value in Eq. (11).
This observation underlines the importance of effective vertices
for the coupling of the strange quarkonium to the nucleon
(compare \cite{Kir95} for previous work).

Our result can be reformulated in terms of an effective $\eta NN$ Lagrangian
with PS-PV mixing \cite{Gro90} that was also discussed in eta photoproduction
before \cite{Ben95}
\begin{equation}
{\cal L}_{\eta NN}(x)=-i g_\eta \bar\Psi(x)[\lambda\gamma_5\phi_\eta + 
   (1-\lambda)\frac{1}{2m_N}i\gamma_\mu \gamma_5 \partial^\mu \phi_\eta]
                            \Psi(x)\, 
\end{equation}
with $\lambda$ being a mixing parameter between the two limiting cases
of PV coupling ($\lambda=0$) and PS coupling ($\lambda=1$).
In combination with Eqs. (26-27) we obtain
\begin{eqnarray}
 \lambda &=& \left( 1+\frac{2m_N}{m_\eta}\frac{f_{\eta NN}}{g_{\eta NN}}
 \right)^{-1} = 0.504 \\
 g_\eta &=&  \frac{g_{\eta NN}}{\lambda} = \frac{f_{\eta NN}}{1-\lambda}
 = 4.03 \\
 \frac{g_{\eta}^2}{4 \pi} &=& 1.29 \,.
\end{eqnarray}
In Fig. 3 we show a calculation of eta photoproduction using our coupling 
constants in comparison with the experimental data. We also compare with
the results of Ref. \cite{Tia94} obtained in PS coupling with their best-fit
coupling constant of $g_{\eta NN}^2 /4\pi =0.4$.
The average result over the angular distribution and, consequently, the total
cross section is about the same in both
calculations, however, in the forward-backward asymmetry our present
calculation provides an even better description due to the small PV admixture.
Considering the dash-dotted lines, calculated using a large value for 
the $g_{a_0NN}$ coupling constant, it becomes clear that such large values
for the $a_0$ coupling and consequently for the $\eta$ coupling are ruled
out by the experiment. 

Our considerations show that towards a better
understanding of the $\eta N$ coupling 
precise measurements
of the $a_0(980)$ decay properties 
as well as a better knowledge on the $a_0NN$ coupling constant
are needed.

We arrive at the conclusion
that both the pseudoscalar and
pseudovector coupling constants
of the $\eta $ meson to the nucleon
seem to be exhausted by the effective
$a_0\pi N$ triangular vertex. Consequently,
the meson cloud model predicts
realistic results for reactions involving 
the coupling of the $s\bar s$ system to nucleons.

To summarize, we wish to stress that in calculating the $\eta NN$
coupling it is necessary to account for the principal difference between the
isosinglet and isotriplet axial vector currents of the nucleon on the quark 
level, a fact ignored by the quark model.
For this reason the three flavor constituent quark model is unable to predict
the correct size for $g_{\eta NN}$.

A similar situation is observed for the case of the $KN\Lambda$- and 
$KN\Sigma$-couplings which are concluded from photoproduction data on the
nucleon to be about an order of magnitude smaller than the quark model
predictions \cite{Mart95}. The small value for $g_{KN\Lambda}$ is well
understood in accounting for the principal difference between the strangeness
preserving and strangeness changing axial vector currents of the nucleon
on the quark level \cite{Kir96}.

\vspace{1cm}
\noindent
{\bf Acknowledgements}

This work was partly supported by the Deutsche Forschungsgemeinschaft
(SFB 201).

\newpage
\noindent
{\bf Figure captions}
\vspace{0.5cm}

Fig. 1\hspace{0.2cm} Axial current dominance (''toy'') model for
the PV coupling of pseudoscalar non strange mesons to the nucleon.
Here $l^\mu$ is the external axial current,
$J_M^{\mu (\alpha ,s)}=iq^\mu\phi_M^{(\alpha , s)}$ denotes
the respective
isovector (upper index $\alpha$) or isosinglet
(upper index $s$) axial current of the $M=\pi , \eta, \eta^\prime$ meson
whereas
$A_\mu^\alpha = g_A \bar u \gamma_\mu\gamma_5 {\tau^\alpha\over 2 }u$
and $A_\mu^s =G_1^s \bar u \gamma_\mu\gamma_5 u$ 
in turn stand for
the isovector and isosinglet axial vector currents
of the nucleon. \\

Fig. 2 \hspace{0.2cm} The effective $\pi a_0 N $ triangular
$\eta NN $ vertex.
The full fat line denotes the $\eta $ meson while dashed and double
lines have been used for the $\pi $ and $a_0$ mesons, respectively.\\

Fig. 3 \hspace{0.2cm} Differential cross section for eta photoproduction
on the proton at different photon lab. energies calculated with the model
of Tiator, Bennhold and Kamalov \cite{Tia94}. The full lines are calculated
with the coupling constants of Eqs. (26-27) and the dash-dotted
lines use coupling constants that were scaled up by the larger value
of 5.79 for $g_{a_0 NN}$, resulting in $g_{\eta NN}=3.78$ and
$f_{\eta NN}=1.08$. The dotted lines show the results of Ref.
\cite{Tia94} in a pure PS model with $g_{\eta NN}^2/4\pi = 0.4$.
The experimental data are from Krusche et al \cite{Kru95}.

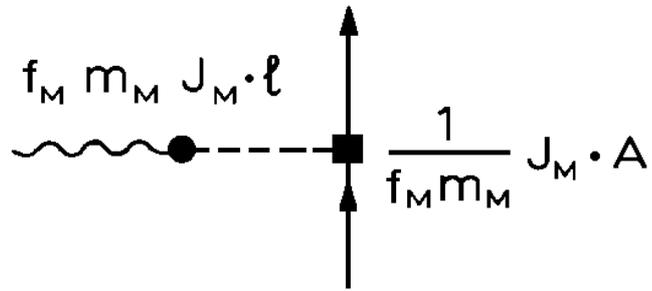
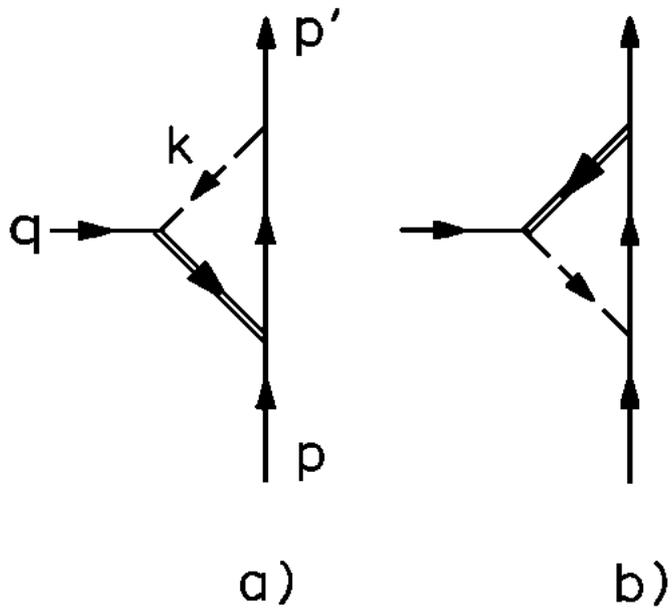
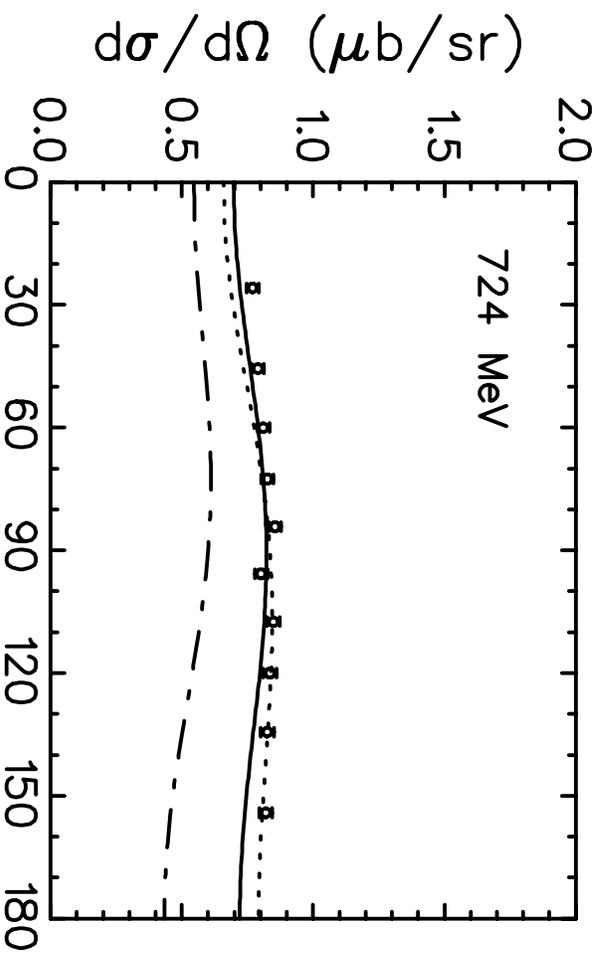
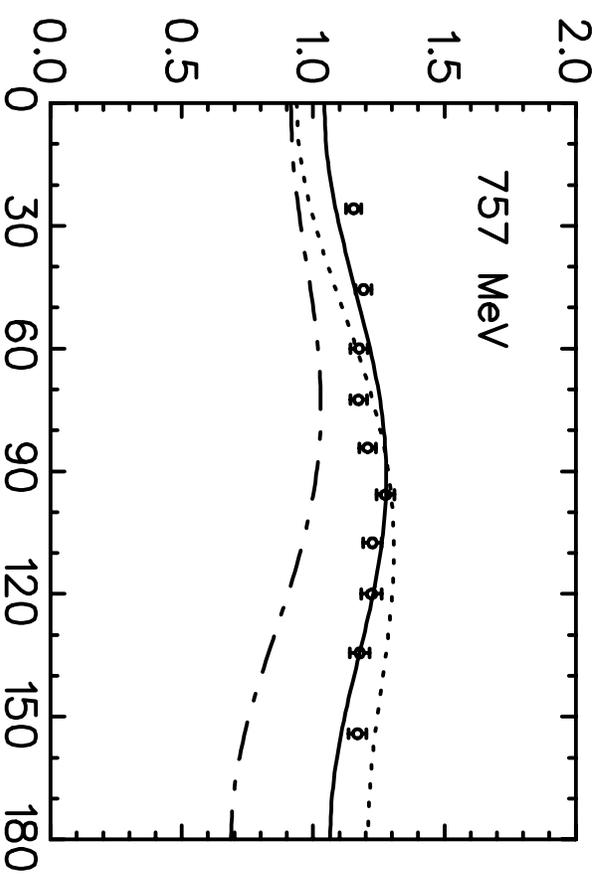
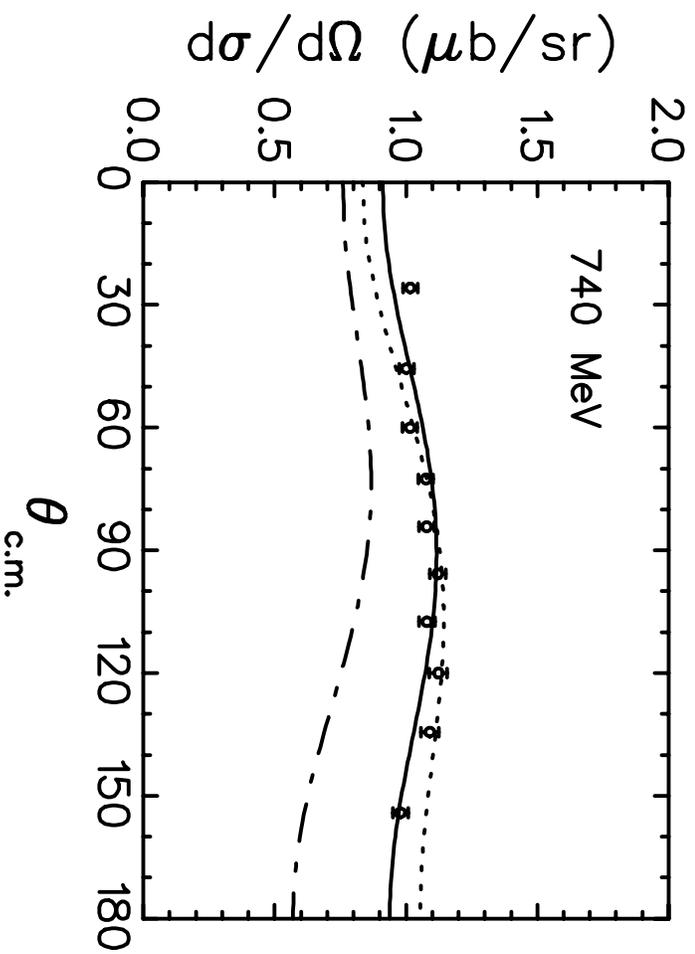
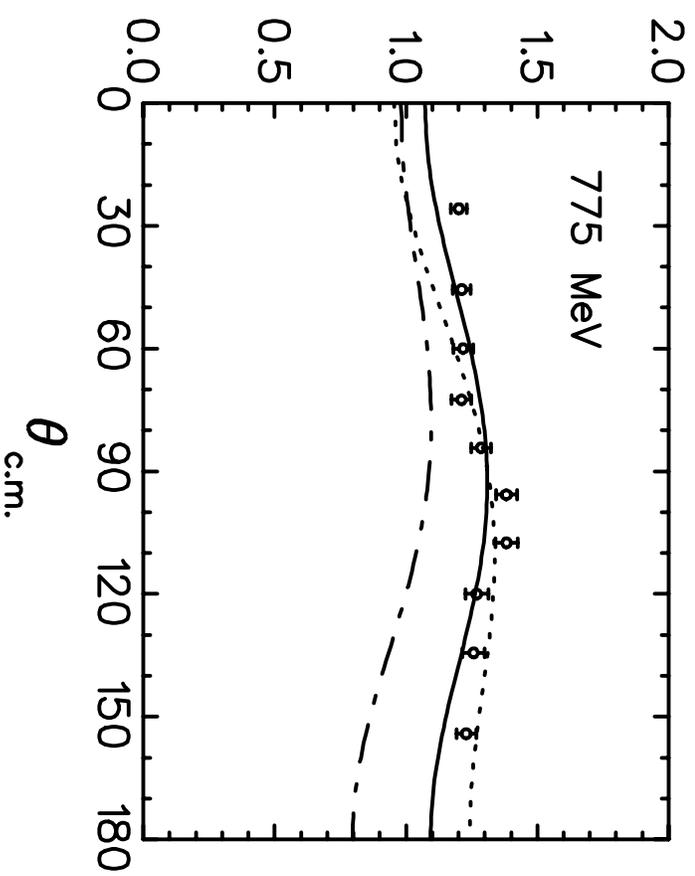
\end{document}